\documentclass[11pt]{article}
\def\Xe{x_{\epsilon}}
\begin{document}

\begin{titlepage}
\begin{flushright}
NSF-KITP-03-95\\
ITEP-TH-55/03
\end{flushright}

\begin{center}
{\Large $ $ \\ $ $ \\
Speeding Strings}\\
\bigskip\bigskip
{\large Andrei Mikhailov\footnote{e-mail: andrei@kitp.ucsb.edu}}
\\
\bigskip\bigskip
{\it Kavli Institute for Theoretical Physics, University of California\\
Santa Barbara, CA 93106, USA\\
\bigskip
and\\
\bigskip
Institute for Theoretical and 
Experimental Physics, \\
117259, Bol. Cheremushkinskaya, 25, 
Moscow, Russia}\\

\vskip 1cm
\end{center}

\begin{abstract}
There is a class of single trace operators in ${\cal N}=4$ Yang-Mills 
theory  which are related by the AdS/CFT correspondence to classical string 
solutions. Interesting examples of such solutions corresponding
to periodic trajectories of the Neumann  system  were studied
recently. In our paper we study a generalization of these solutions. 
We consider strings moving with  large velocities. 
We show that the worldsheet of the fast moving string can be considered 
as a perturbation of the degenerate worldsheet, with the small parameter 
being the relativistic factor $\sqrt{1-v^2}$. The series expansion in this
relativistic factor should correspond to the perturbative expansion
in the dual Yang-Mills theory. The operators minimizing the anomalous
dimension in the sector with given charges correspond to periodic
trajectories in the mechanical system which is closely related
to the product of two Neumann systems. 
\end{abstract}

\end{titlepage}

\section{Introduction.}
The operators with the large spin or large R charge in ${\cal N}=4$
Yang-Mills theory are of a special interest for  the AdS/CFT correspondence. 
The study of these operators initiated in \cite{BFFHP,Metsaev,BMN,GKP}
gives  the most convincing evidence for the validity of 
the Maldacena conjecture. The R charge of the operator 
is specified by the representation of the R-symmetry algebra, or equivalently
 by the highest weight $[J_1,J_2,J_3]$. It was shown in \cite{BMN} that  the
operators with $J_1\gg J_2, J_3$ (now known as BMN operators) 
have anomalous dimensions which can be
represented as power series in $\lambda/ J_1^2$ where $\lambda$ is the
t'Hooft coupling constant.  The authors of \cite{BMN} derived this
series expansion from the string theory, and later it was reproduced
in the Yang-Mills perturbation theory. Recently more general operators
with $J_1, J_2$ and $J_3$ of the same order of magnitude and also with
large spins were studied
in the series of papers 
\cite{FT02,Tseytlin,Russo,FT03,FT03Q,FT03R,AFRT,MZ,BS,BMSZ,BFST,EMZ}. 
Unlike the BMN operators, these operators  
cannot be considered as small deformations of the BPS operators. 
But their anomalous dimension   is of the order $\lambda/J$,
and there is a conjecture
 that in the limit of large  charges one can compute it
from the energy of the corresponding string theory state
in the large radius $AdS_5\times S^5$. 

It turns out that the corresponding string theory state is a single
semiclassical string. The simplest case to consider is  a spin zero operator. 
The worldsheet of the corresponding string is a product
of a timelike geodesic in $AdS_5$ and a rotating contour in $S^5$. 
The authors of \cite{AFRT} suggested the following ansatz for this solution:
\begin{equation}
x_1+ix_2=x_1(\sigma) e^{i{w_1\over k}t},\;\;
x_3+ix_4=x_2(\sigma) e^{i{w_2\over k}t},\;\;
x_5+ix_6=x_3(\sigma) e^{i{w_3\over k}t}
\end{equation}
Here $x_1,\ldots, x_6$ are the coordinates on a sphere $S^5$ subject
to the constraint $x_1^2+\ldots +x_6^2=1$, and $t$ is a length
parameter on the timelike geodesic in $AdS_5$. Substitution of this ansatz 
into the string worldsheet action leads to the one-dimensional mechanical 
system with the Lagrangian:
\begin{equation}
 L={1\over 2}\sum\limits_{i=1}^3 ((\partial_{\sigma} x_i)^2-w_i^2x_i^2)
\end{equation}
where $x_i$ are restricted to  a sphere: $\sum\limits_{i=1}^3 x_i^2=1$.
The corresponding R-charges are
\begin{equation}
J_i=\sqrt{\lambda}w_i\int {d\sigma\over 2\pi} x_i^2(\sigma)
\end{equation}
For comparison to the field theory computation the most interesting
case is when the momenta $J_i$ are very large. 
It was noticed in \cite{MateosMateosTownsend} that this
limit corresponds to the string moving very fast, and  the 
induced metric on the string worldsheet becoming nearly degenerate.
In some sense the Yang-Mills perturbation series for such operators
 should correspond to the expansion in the powers
of the relativistic factor $\sqrt{1-v^2/c^2}$. 

 Motivated by this
observation we study in this paper  
classical strings in $AdS_5\times S^5$ moving with large  
velocities. Such solutions correspond to the Yang-Mills operators
with large charges and spins. 
We will be interested in the operators which 
have  very large charges $J_a$
and finite but small ratios $\lambda/J_a^2$.
We will pick a special combination of charges which we will denote
 $Q_{tot}$ and define
\begin{equation}
\epsilon^2={\lambda\over Q_{tot}^2}
\end{equation}
Suppose that we have a uniform definition of
the operator for all  values of the coupling constant $\lambda$, 
and the operator corresponds to the classical
string solution in $AdS_5\times S^5$. When we change the coupling constant
the shape of the worldsheet changes. Of course, the radius of
$AdS_5\times S^5$ also changes as $\lambda^{1/4}$, but 
this is just the overall coefficient in front of the metric.
Given a classical string solution in $AdS_5\times S^5$ we actually get
 the whole one-parameter family of  classical solutions 
by varying the coupling constant $\lambda$. We will call
this family of string worldsheets $\lambda$-{\it family}.
In the regime we are interested in it is  convenient to parametrize
this $\lambda$-family by $\epsilon$. 
Let $\Sigma(\epsilon)$ be the worldsheet
of the string corresponding to our  operator at the coupling constant
$\lambda=Q_{tot}^2\epsilon^2$. We will be interested in the class of
operators such that the corresponding worldsheet $\Sigma(\epsilon)$
has a well-defined limit when $\epsilon\to 0$.
In this limit it becomes a null-surface $\Sigma(0)$, a surface
with the degenerate metric ruled by the light rays. Moreover, 
this null-surface 
naturally comes with an additional structure. 
This additional structure
 is a function $\sigma:\Sigma(0)\to S^1$ constant
along the light rays. 
In other words, if we think of
$\Sigma(0)$ as a collection of the light rays,
then this collection is a one-parameter family, parametrized by
$\sigma$.  This function $\sigma$ is determined by the shape of
$\Sigma(\epsilon)$ for very small but nonzero $\epsilon$.
It is defined modulo the overall shift (if 
$\sigma_1=\sigma_2+$const, then $\sigma_1$ and $\sigma_2$
should be considered equivalent.) The definition requires the
choice of a particular combination of symmetries (corresponding
to the charge $Q_{tot}$.) In fact
${\sqrt{\lambda}\over\epsilon}d\sigma$ is the density
of the charge $Q_{tot}$ on the worldsheet $\Sigma(\epsilon)$ in
the limit $\epsilon\to 0$.

We conjecture that the $\lambda$-family  (and the corresponding YM operator) 
is in fact uniquely determined by the null-surface $\Sigma(0)$ and 
the real function $\sigma$.  
Indeed, to specify the $\lambda$-family it should be enough to
specify the worldsheet at some finite value of $\lambda$. 
The worldsheet is parametrized by 16 real functions of one real variable,
8 functions specifying the shape of the string at time zero
and 8 functions specifying the velocities. But $\Sigma(0)$ is
specified by 15 real functions, plus one real function $\sigma$,
therefore we get also 16 real functions. This counting of the
parameters leads us to the conjecture that there is a correspondence
between the non-degenerate extremal surfaces and the degenerate
surfaces with a function $\sigma$ constant on the light rays.
We must stress that this construction uses the assumption that
there is a uniform definition of the Yang-Mills operator
for all values of the coupling constant $\lambda$. 

Let us summarize. The Yang-Mills operators
(of certain type) are in one-to-one correspondence
with the pairs $(\Sigma(0),\sigma)$ where $\Sigma(0)$ is a null-surface
in $AdS_5\times S^5$ and $\sigma$ is a function from
$\Sigma(0)$ to $S^1$ constant on the light rays. On the other
hand, the operator determines a family of contours $\Sigma(\epsilon)$
where $\epsilon^2=\lambda/J^2$,  such that
the limit when $\epsilon\to 0$ is $\Sigma(0)$.
This means that for each null-surface $\Sigma(0)$ with a function
$\sigma$ there should be a preferred family of deformations
$\Sigma(\epsilon)$.

In this paper we will focus on a special class of operators 
for which a particular combination of their charges 
and their conformal dimension is small (compared to
other charges.) 
In perturbation theory such operators are defined
by the requirement that their engineering dimension
is equal to certain linear combination of their spins and
 R-charges. The combination of the dimension and charges
which is small 
corresponds in the AdS picture 
to certain lightlike Killing vector $V$ 
of $AdS_5\times S^5$. 
In the limit of infinite velocities the worldsheet becomes
a null-surface spanned by the integral curves of $V$ 
(which are automatically null-geodesics.) A null surface spanned
by the integral curves of $V$ is specified by a curve in the
coset space $(AdS_5\times S^5)/V$ such that the tangent direction to the
curve is orthogonal to $V$. Moreover,  we have a function
$\sigma$ constant on the light rays which we can use to
parametrize the curve.
Therefore we get a map $C: \; S^1\to (AdS_5\times S^5)/V$.
The charge corresponding to $V$ is given in the first order
in $\epsilon$ by the "action functional"
\begin{equation}\label{CosetAction}
\epsilon \sqrt{\lambda}\int_{S^1} d\sigma \; g_{ij}(C) \;
{d C^i(\sigma)\over d\sigma} {d C^j(\sigma)\over d\sigma}
\end{equation}
where $g_{ij}$ is the metric on $(AdS_5\times S^5)/V$. This charge 
corresponds to the one loop anomalous dimension of the operator.
The Killing vector fields other than $V$ correspond to the
charges which go to infinity in the limit $\epsilon\to 0$.
These "large" charges are of the form 
${\sqrt{\lambda}\over\epsilon}\int d\sigma F(C)$ with certain function $F$.
The particular combination of charges which we denoted $Q_{tot}$ corresponds
to  $F(C)\equiv 1$. 

Therefore the operator defines formally
a closed trajectory of the particle moving in $(AdS_5\times S^5)/V$ 
under the constraint that the velocity is orthogonal to $V$. 
The action on this trajectory corresponds to the one-loop anomalous dimension
of the operator. The trajectory does not have to satisfy any
equations of motion.

Perhaps one could define a measure on the space of contours in
$(AdS_5\times S^5)/V$, which would compute how many operators
 have the anomalous dimension in a given interval.
One  concrete question  is
what is the minimal one-loop anomalous dimension in the sector
with the given charges. At the level of
one loop this problem reduces to finding periodic
trajectories of certain C.~Neumann type mechanical system which
is a natural generalization of the one considered in
\cite{AFRT}. We explain the details in Section 3. The Neumann
systems are closely related to integrable spin chains
--- see the discussion in \cite{Gorsky}.

It would be interesting to understand in general when
two string worldsheets belong to the same $\lambda$-family. 
For the operators which we are discussing 
in this paper the dependence of the string worldsheet on $\lambda$
is of the crucial importance. Indeed, the string worldsheet action 
is proportional to  $R^2/\alpha'\sim \sqrt{\lambda}$, 
and because of that one could
naively guess that the anomalous dimension of the operator is of the
order $\sqrt{\lambda}$ in the strong coupling regime. 
But this is wrong in the case we are considering 
precisely because the shape of the string worldsheet itself depends
on $\lambda$, and in such a way that the $V$-charge of
the $\lambda$-dependent worldsheet is proportional to 
$\lambda$ rather than $\sqrt{\lambda}$. 

If we want to know the anomalous dimension of the operator
as a function of $\lambda$, it is not enough to know
just the string worldsheet corresponding to this operator.
We have to know the whole $\lambda$-family.
But there are some questions which can be answered without
the knowledge of the $\lambda$-families. We can ask what
is the minimal anomalous dimension of the operator in the
sector with the given charges, for the given value
of the coupling constant. In perturbation theory, the answer
will be a series in $\lambda$.
If we want to answer this question in the strong
coupling regime using the AdS/CFT correspondence we have to find
the string worldsheet which has a minimal energy for given
charges. For different values of the coupling (different
radii of the AdS space) we will get different worldsheets.
Therefore, we will get some family of worldsheets --- but
it is not guaranteed that this will be a $\lambda$-family!
Indeed, there is no apriori reason why the same operator
would minimize the anomalous dimension in the sector
with the given charges, for different values of the coupling
constant.

\vspace{5pt}

{\em The plan of the paper.}
In Section 2 we will give a definition of
the null-surface and explain that the null surface is the limit of
the family of 
ultrarelativistic extremal surfaces. We will also explain how to compute
the conserved charges in this limit. 
In Section 3 we will study the ultrarelativistic surfaces in $AdS_5\times S^5$.
We will show that the operators minimizing the one loop anomalous dimension
for given charges correspond  to  the periodic trajectories of 
the gauged Neumann system. We briefly 
review the basic properties of the Neumann systems.
We explain that the space of periodic trajectories consists of several
branches corresponding to the possible degenerations of the invariant tori.

After this paper was completed we have received the preprint
\cite{ArutyunovRussoTseytlin} which has an overlap with our 
paper. 

\vspace{5pt}

{\em Note added in the revised version.} 
We have made a mistake in the original version
of this paper which lead us to the conclusion that the correspondence
between the operators and the pairs $(\Sigma(0),\sigma)$ can not be
one-to-one. We claimed that in order to specify the $\lambda$-family,
one has to know besides $(\Sigma(0),\sigma)$ also 
$\Sigma(\epsilon)$ to the first order in $\epsilon$. We realized
that we made a mistake studying the recent preprint \cite{Kruczenski}
which contains the construction of $\Sigma(\epsilon)$ to the
first order in $\epsilon$ from the known $\Sigma(0)$ and $\sigma$,
for $\Sigma(0)$ generated by the orbits of $V$.

We have also learned about the papers \cite{AleksandrovaBozhilov}
where the degenerate surfaces were discussed
in the context of Frolov-Tseytlin solutions. The authors
of \cite{AleksandrovaBozhilov} studied  ultrarelativistic
strings in the backgrounds with the $B$-field, as well as
ultrarelativistic membranes.

The null-surface perturbation theory was considered
in a closely related context in \cite{dVGN}.

\section{Null surfaces and their deformations.}
\subsection{Definition of a null surface.}
Consider a two-dimensional surface $\Xi$ embedded in the
metric space $M$ with Minkowski signature $(1,d-1)$, $d>2$. 
We will say that $\Xi$
is {\it isotropic} if the induced metric is degenerate. This means
that for every point $x\in\Xi$ the tangent space $T_x\Xi$
has a null-vector $v(x)$ which is orthogonal to all other vectors
in $T_x\Xi$. The vectors in $T_x\Xi$ which are not parallel
to $v(x)$ are all space-like. The integral curves of the vector
field $v(x)$ will be called null-curves.

We will call an isotropic surface $\Xi$ a {\it null-surface} if
its null-curves are light rays (null geodesics) in $M$.

For example, let us consider a future light cone of a point $x_0\in M$.
It is generated by the light rays emitted at $x_0$. If we choose
any one-parameter family of the light rays emitted at $x_0$, then
this family will sweep a null surface. 
More general examples can be obtained in the following way.
Consider a spacelike curve $C$. For each point  $x\in C$
pick a light ray going through $x$, such that the direction
of this light ray at the point $x$ is orthogonal to $T_xC$.
The resulting one-parameter family of light rays will be
a null surface. 

We are interested in null-surfaces because they can be thought of
as worldsheets of ultrarelativistic strings. The classical 
equations of motion 
imply that the string worldsheet is an extremal surface. 
If the string moves very fast the surface becomes isotropic.
Moreover, one can see that the isotropic limit of the extremal
surface should be a null surface. Indeed, let us
introduce on the string worldsheet the coordinates $\xi^+$,
$\xi^-$, so that the induced metric is $\rho(\xi^+,\xi^-)d\xi^+ d\xi^-$. 
The condition that the surface is extremal is
\begin{equation}\label{Extremal}
{D\over \partial \xi^+}{\partial x^{\mu}\over\partial \xi^-}=0
\end{equation}
where $x^{\mu}(\xi^+,\xi^-)$ are the embedding functions.
In the limit when the surface becomes isotropic, the two null-directions
${\partial\over\partial \xi^+}$ and ${\partial\over\partial\xi^-}$
coincide, and Eq. (\ref{Extremal}) implies that the limiting null-curves
are geodesics. We will argue that any null-surface
can be obtained as a limit of a family of extremal surfaces.

\subsection{Ultrarelativistic surfaces.}
We are interested in the case when the space-time admits
 a light-like Killing
vector field $V$. We will require
that the light rays forming the null-surface $\Sigma(0)$
are the integral curves of $V$. Consider a family 
of surfaces $\Sigma(\epsilon)$ which approach $\Sigma(0)$
when $\epsilon\to 0$, in the sense that the coordinate difference
between $\Sigma(0)$ and $\Sigma(\epsilon)$ is of the order
$\epsilon^2$.
We will now describe a special choice of coordinates on 
$\Sigma(\epsilon)$.

Suppose that $\Sigma(\epsilon)$ is topologically a cylinder.
Pick a closed space-like contour $C_0\subset \Sigma(\epsilon)$
For each point  $x\in C_0$ take a direction in $T_x\Sigma(\epsilon)$
which is orthogonal to $T_xC_0\subset T_x\Sigma(\epsilon)$. Since 
$\Sigma(\epsilon)$ is close to $\Sigma(0)$ this direction
is close to the direction of $V$. Therefore we can choose
a vector $u(x)\in T_x\Sigma(\epsilon)$ which is orthogonal to 
$T_xC_0$ and $u(x)=V(x)+O(\epsilon^2)$. We have 
$(u(x),u(x))\simeq\epsilon^2$. Let us introduce the parametrization 
$\sigma$ of $C_0$ in the following way:
\begin{equation}
\left({\partial x(\sigma)\over\partial\sigma},
{\partial x(\sigma)\over\partial\sigma}\right)=
-{1\over \epsilon^2} (u(x),u(x))
\end{equation}
The coordinates $\tau$ and $\sigma$ on
$\Sigma(0)$ are determined by the conditions on the embedding
functions $x(\tau,\sigma)$:
\begin{eqnarray}
&& (\partial_{\tau}x,\partial_{\tau}x)=-\epsilon^2
(\partial_{\sigma}x,\partial_{\sigma}x),\;\;\;\;
(\partial_{\tau}x,\partial_{\sigma}x)=0\\[5pt]
&&x(0,\sigma)=x(\sigma)\in C_0
\end{eqnarray}

The first equation uniquely determines 
$\partial_{\tau}x(\tau_0,\sigma)
\in T_{x(\tau_0,\sigma)}\Sigma(\epsilon)$
from the contour $x(\tau_0,\sigma)$ for a fixed $\tau=\tau_0$.
Therefore it is an "evolution equation" for the contour on
the worldsheet.
The second equation determines the initial condition --- the
contour at $\tau=0$. The equation for the extremal surface
in these coordinates is 
$D_{\tau}\partial_{\tau} x=\epsilon^2 D_{\sigma}\partial_{\sigma} x$. 
It implies that at the zeroth
order in $\epsilon$, $x(\tau,\sigma_0)$ for a fixed $\sigma_0$
is a  null geodesic. Therefore $\partial_{\tau}x=V+O(\epsilon^2)$
not only on the initial curve $C_0$ but everywhere on the worldsheet.

To summarize, we have chosen the coordinates $\sigma,\tau$ on
$\Sigma(\epsilon)$ such that the embedding functions 
$x_{\epsilon}(\sigma,\tau)$
satisfy the constraints
\begin{eqnarray}\label{Constraints}
&&\left({\partial \Xe\over \partial \tau},
{\partial \Xe\over \partial\tau}\right)+
\epsilon^2 \left({\partial \Xe\over\partial\sigma},
{\partial\Xe\over\partial\sigma}\right)=0\nonumber\\[5pt]
&&\left(
{\partial \Xe\over \partial \tau},
{\partial \Xe\over \partial\sigma}\right)=0 \label{SigmaOrthogonalTau}
\end{eqnarray}
and the equations of motion
\begin{equation}\label{EqM}
{1\over\epsilon}D_{\tau}\partial_{\tau}\Xe-
\epsilon D_{\sigma}\partial_{\sigma}\Xe=0
\end{equation}
 There is a residual gauge 
invariance; the constraints (\ref{Constraints}) and the equations
(\ref{EqM}) are preserved by the infinitesimal vector fields
\begin{equation}\label{Residual}
\left[f_L(\sigma+2\epsilon\tau)+f_R(\sigma-2\epsilon\tau)\right]
{\partial\over\partial\tau}
+\epsilon\left[f_L(\sigma+2\epsilon\tau)-f_R(\sigma-2\epsilon\tau)
\right]{\partial\over\partial\sigma}
\end{equation}
We want to study the solutions to the equations (\ref{EqM}) 
and the constraints
(\ref{Constraints}). Let us consider the following ansatz:
\begin{equation}\label{Ansatz}
\Xe(\sigma,\tau)=x_0(\sigma,\tau)+
\epsilon^2\eta_1(\sigma,\tau)+\epsilon^4\eta_2(\sigma,\tau)+\ldots
\end{equation}
This ansatz is preserved by the vector fields (\ref{Residual}) with
$f_L=f_R$.
The equations of motion and constraints for $x$ imply:
\begin{eqnarray}
&&{D^2\eta_1 \over \partial\tau^2}
+R\left({\partial x_0\over\partial\tau},\eta_1\right)
{\partial x_0\over\partial\tau} =
{D\over\partial\sigma}{\partial x_0\over\partial \sigma}
\\[5pt]
&&(D_{\sigma}\eta_1 , \partial_{\tau}x_0)+
(D_{\tau}\eta_1 ,\partial_{\sigma}x_0)=0\label{FirstEtaConstraint}
\\[5pt]
&&(D_{\tau}\eta_1 ,\partial_{\tau}x_0)=
-{1\over 2}(\partial_{\sigma}x_0,
\partial_{\sigma}x_0)\label{SecondEtaConstraint}
\end{eqnarray}
where 
$R(\xi,\eta)=-\nabla_{\xi}\nabla_{\eta}+
\nabla_{\eta}\nabla_{\xi}
+\nabla_{[\xi,\eta]}
$
is the curvature tensor.
The first equation detremines $\eta_1(\sigma,\tau)$ from the initial
conditions $\eta_1(\sigma,0)$ and 
$\partial_{\tau}|_{\tau=0}\eta_1(\sigma,\tau)$ which 
should satisfy the constraints (\ref{FirstEtaConstraint}) and
(\ref{SecondEtaConstraint}).
The higher coefficients $\eta_i$ satisfy  the similar
equation and constraints. But as we have explained in the Introduction, we 
expect that $\eta_i$ are actually determined from $x_0$.
This is because we want the solutions for different $\epsilon$ to 
correspond to the same operator in the gauge theory but at the different
values of the coupling constant. For the same reason we have omitted
in our ansatz (\ref{Ansatz}) the odd powers of $\epsilon$. The parameter
$\epsilon$ is proportional to $\sqrt{\lambda}$;
we hope that the string solution has an expansion in integer powers
of $\lambda$.

\subsection{Conserved quantities.}
The string worldsheet action is
\begin{equation}
S={\sqrt{\lambda}\over {4\pi}}
\int d\sigma d\tau \left[
{1\over\epsilon}(\partial_{\tau}x,\partial_{\tau}x)
-\epsilon (\partial_{\sigma}x,\partial_{\sigma}x)
\right]
\end{equation} 
Suppose that the spacetime admits the Killing vector field $W^{\mu}(x)$.
The corresponding conserved charge is given by the integral over
the spacial slice of the string worldsheet of the closed form
\begin{equation}
*j={\sqrt{\lambda}\over {4\pi}}
\left(
{1\over\epsilon}W_{\mu}(x)\partial_{\tau}x^{\mu}d\sigma
+\epsilon W_{\mu}(x)\partial_{\sigma}x^{\mu}d\tau\right)
\end{equation}
On the ansatz (\ref{Ansatz}):
\begin{eqnarray}
&&*j={\sqrt{\lambda}\over {4\pi}}\left[
{1\over\epsilon}W_{\mu}(x_0)\partial_{\tau}x_0^{\mu}\;d\sigma
+
\right.\nonumber\\[5pt]&&\left.+
\epsilon 
\left(
\eta_1^{\nu}\nabla_{\nu}W_{\mu}(x_0)\;\partial_{\tau}x_0^{\mu}\; d\sigma+
W_{\mu}(x_0)D_{\tau}\eta_1^{\mu}\; d\sigma
+W_{\mu}(x_0)\partial_{\sigma}x_0^{\mu}\; d\tau\right)
+\right.\nonumber \\[5pt]
&&\left. +O(\epsilon^3)\right]
\end{eqnarray}
Let us consider two conserved charges, the charge corresponding to
$V$ and the charge corresponding to some other Killing vector
field $U\neq V$.
The charge corresponding to $U$ is
\begin{equation}\label{QU}
Q_U={1\over \epsilon}{\sqrt{\lambda}\over {4\pi}}\int
d\sigma U_{\mu}
\partial_{\tau}x^{\mu}_0+O(\epsilon)
={1\over \epsilon}{\sqrt{\lambda}\over {4\pi}}\int
d\sigma\; U_{\mu}V^{\mu}
+O(\epsilon)
\end{equation}
and the charge corresponding to $V$ is
\begin{equation}\label{QV}
Q_V=\epsilon{\sqrt{\lambda}\over {4\pi}}\int d\sigma\;
\partial_{\tau}x_0^{\mu}D_{\tau}\eta_{1\mu}
=-\epsilon{\sqrt{\lambda}\over {8\pi}}\int d\sigma\; 
\partial_{\sigma}x_{\mu}\partial_{\sigma}x^{\mu}+O(\epsilon^3)
\end{equation}
where we have taken into account the constraint 
(\ref{SecondEtaConstraint}).
From Eq. (\ref{QU}) 
\begin{eqnarray}
&&\epsilon={\sqrt{\lambda}\over 4\pi Q_U}\left[\int
d\sigma \; U_{\mu}V^{\mu}+O({\epsilon^2})\right]=\\[5pt]
&&={\sqrt{\lambda}\over 4\pi Q_U}\left[\int
d\sigma \; U_{\mu}V^{\mu}+O\left({\lambda\over Q_U^2}\right)\right]
\end{eqnarray}
Then (\ref{QV}) implies that in the limit $Q_U\to \infty$
\begin{equation}\label{VChargeGeneralCase}
Q_V={1\over 32\pi^2}{\lambda\over Q_U}\left[
\int d\sigma\;U_{\mu}V^{\mu}
\int d\sigma\;(\partial_{\sigma}x_0,\partial_{\sigma}x_0)+\ldots
\right]
\end{equation}
where dots stand for the subleading terms. If both $Q_U$ and $\lambda$
go to infinity in such a way that $\lambda\over Q_U^2$ is finite
but small, then these subleading terms are the power 
series in $\lambda\over Q_U^2$. The coefficients of these power
series explicitly depend on  
$\eta_k (\sigma,\tau)$.

Let us summarize what we have.
At the first order in $\lambda$ the V-charge 
(\ref{VChargeGeneralCase}) is proportional to the
``action functional'' for the trajectory $x_0(\sigma)$
which determines the degenerate surface $\Sigma(0)$:
\begin{equation}
S=\int d\sigma (\partial_{\sigma}x_0,\partial_{\sigma}x_0)
\end{equation}
For $\Sigma(0)$ to be degenerate, 
the velocity is constrained to be orthogonal to $V$:
\begin{equation}
(V,\partial_{\sigma}x_0)=0
\end{equation}
And we have to remember that two contours $x_0(\sigma)$
which are different by the $\sigma$-dependent shift along
$V$ give the same surface $\Sigma(0)$, therefore the 
``gauge symmetry'':
\begin{equation}
\delta x_0^{\mu}(\sigma)= \delta\phi(\sigma) \; 
V^{\mu}(x_0(\sigma))
\end{equation}
where $\delta\phi(\sigma)$ is a $\sigma$-dependent parameter.

\section{Ultrarelativistic surfaces in $AdS_5\times S^5$.}
\subsection{Relations between the charges in the ultrarelativistic limit.}
The space $AdS_5\times S^5$ is the product of a hyperboloid in ${\bf R}^{2+4}$ 
and a sphere in ${\bf R}^6$. We will choose a complex structure in 
${\bf R}^{2+4}$ and ${\bf R}^6$ and introduce the complex coordinates
$Y_I$ , $I=0,1,2$ in ${\bf R}^{2+4}$ and $Z_I$, $I=1,2,3$ in ${\bf R}^6$.
The hyperboloid and the sphere
are given by the equations $|Y_0|^2-|Y_1|^2-|Y_2|^2=1$ and
$|Z_1|^2+|Z_2|^2+|Z_3|^2=1$ respectively. We will denote $Y_I$ and $Z_I$
collectively as $X_A$, $A=0,1,\ldots,5$. We will put 
$(X_0,X_1,X_2)=(Y_0,Y_1,Y_2)$
and $(X_3,X_4,X_5)=(Z_1,Z_2,Z_3)$. Let us also define the sign factor
$s_I$, $(s_0,s_1,s_2)=(1,-1,-1)$. 

The symmetry group of $AdS_5\times S^5$ is $SO(2,4)\times SO(6)$.
The Cartan torus is six-dimensional. We can choose it to be represented by
the six Killing vector fields $U_0, \ldots, U_5$:
\begin{equation}
U_A. X_B= i\delta_{AB} X_B
\end{equation}
We will choose the light-like Killing vector field to be 
$V=\sum\limits_{A=0}^5 U_A$. The corresponding charges are:
\begin{eqnarray}
&&Q_{U_A}={1\over\epsilon}
{\sqrt{\lambda}\over 4\pi}\int d\sigma\;U_A^{\mu}V_{\mu}
={1\over\epsilon}
{\sqrt{\lambda}\over 4\pi}\int d\sigma\;s_A |X_A|^2 +\ldots
\\[5pt]
&&Q_V=-\epsilon{\sqrt{\lambda}\over {8\pi}}\int d\sigma\; 
\partial_{\sigma}x_{\mu}\partial_{\sigma}x^{\mu}
=\nonumber\\[5pt]
&&=-\epsilon{\sqrt{\lambda}\over {8\pi}}\int_{\tau=\tau_0}d\sigma\; 
\sum_A s_A|\partial_{\sigma}X_A|^2 +\ldots
\end{eqnarray}
where $s_A$ is a sign:
$(s_0,\ldots,s_5)=(-1,1,1,1,1,1)$. Let us introduce a special 
combination of charges:
\begin{equation}
Q_{tot}=Q_{U_0}-Q_{U_1}-Q_{U_2}=Q_{U_3}+Q_{U_4}+Q_{U_5}=
{1\over\epsilon}{\sqrt{\lambda}\over 4\pi}\int d\sigma +\ldots
\end{equation}
Let us require that the period of $\sigma$ is $2\pi$:
\begin{equation}
\int d\sigma=2\pi
\end{equation}
This condition defines $\epsilon$ in terms of 
$\sqrt{\lambda}/Q_{tot}$.
With this notation  
\begin{equation}\label{OneLoop}
Q_V={\lambda\over 16\pi Q_{tot}}
\int d\sigma\;(\partial_{\sigma}x_0,\partial_{\sigma}x_0) +\ldots
\end{equation}

\subsection{Gauged Neumann system.}
Finding the extremum of $Q_V$ for fixed $Q_0,\ldots, Q_5$
is reduced to extremizing the functional
\begin{equation}
S_0[x(\sigma)]=\int d\sigma\;  
\partial_{\sigma}x_{\mu}\partial_{\sigma}x^{\mu}
\end{equation}
for fixed  $q_A=\int d\sigma |X_A|^2$ and with the constraint:
\begin{equation}\label{VOrthogonalDX}
V_{\mu}(X)\partial_{\sigma}X^{\mu}=0
\end{equation}
which follows from Eq. (\ref{SigmaOrthogonalTau}) in the limit 
$\epsilon\to 0$.
The solutions  correspond to periodic 
trajectories of the mechanical system with the following 
Lagrangian:
\begin{equation}
 L=
\sum\limits_{A=0}^5 s_A|\partial_{\sigma} X_A(\sigma)|^2
 - \sum s_A\gamma_A (|X_A(\sigma)|^2-q_A)
 -\Lambda(\sigma)M_{tot}(\sigma)
\end{equation}
where
\begin{equation}
 M_{tot}(\sigma)=i\sum\limits_{A=0}^5 s_A\overline{X}_A(\sigma)
\stackrel{\leftrightarrow}{\partial}_{\sigma} X_A(\sigma)  
\label{MtotConstraint}
\end{equation}
and $X_A$ is restricted to $AdS_5\times S^5\subset {\bf C}^{1+5}$.
The first term is the action of the free particle moving on
$(AdS_5\times S^5)/V$. The other terms are fixing
 $q_A={1\over 2\pi}\int d\sigma |X_A(\sigma)|^2$ and imposing 
the constraint $M_{tot}(\sigma)=0$ 
which follows from (\ref{VOrthogonalDX}).  The constants
 $\gamma_A$ and the function $\Lambda(\sigma)$ 
are the Lagrange multipliers.
The action and the constraint are invariant under
the gauge transformation 
\begin{equation}\label{GaugeTransformation}
X_A(\sigma)\mapsto e^{i\phi(\sigma)}X_A(\sigma),\;\;\;\;
\Lambda(\sigma)\mapsto \Lambda(\sigma) + 
{\partial\phi(\sigma)\over\partial\sigma}
\end{equation}
which corresponds to the residual diffeomorphism invariance
(\ref{Residual}) of the string worldsheet theory. 
One can choose the gauge 
$\Lambda(\sigma)=0$. In this gauge the equations of motion
coincide with the equations of motion of two independent
conventional (not gauged) Neumann systems. Because of the 
constraint and the gauge
symmetry this system has eight degrees of freedom, rather than
ten degrees of freedom as the conventional Neumann system
on $AdS_5\times S^5$ would have.

 Notice that
the kinetic term 
$\sum\limits_{A=0}^5 s_A|\partial_{\sigma} X_A(\sigma)|^2$ is positive 
definite if the constraint $M_{tot}=0$ is satisfied. One can see this
by choosing the gauge $\;$Im$\;X_0=0$. And of course,  the
periodic trajectories we are interested in 
are supposed to be periodic only modulo the gauge transformation.

\subsection{Off-diagonal charges.}
There are 24 Killing vectors in $AdS_5\times S^5$ which do not commute
with $U_0,\ldots, U_5$. The leading expressions for these charges
when $\epsilon\to 0$ are of the form 
$Q_{\mu\nu}=
{1\over\epsilon}\int d\sigma\; x_{\mu}(\sigma)x_{\nu}(\sigma)$.
Suppose that the Killing vector $W$ can be represented
as $W=[V,W']$ where $W'$ is another Killing vector.
Then $(V,W)=0$ and the charge corresponding to $W$ is
 of the order $\epsilon$. But there are symmetries which
cannot be represented as a commutator of another symmetry with $V$.
The corresponding charges are generally speaking of the
order $1\over\epsilon$. The corresponding vector fields are
of the form $W_{\Phi}=I. \mbox{grad}\;\Phi(X)$ where $I$ is the complex
structure and 
$$\Phi(X)=\sum_{I,\bar{J}}
\Phi^{I\bar{J}}Y_I\bar{Y}_{\bar{J}}+\sum_{I\bar{J}}
\tilde{\Phi}^{I\bar{J}} Z_I\bar{Z}_{\bar{J}}$$
and $\Phi,\tilde{\Phi}$ are constant Hermitean matrices
with zeroes on a diagonal. We have $(W_{\Phi},V)=\Phi$. 

Suppose that the  contour is a solution to the Neumann equations
with $A_I\neq A_J$ for $I\neq J$ and $\tilde{A}_K\neq \tilde{A}_L$ for 
$K\neq L$. 
Then the non-diagonal charges are zero in the order $1\over\epsilon$.
Indeed, the coefficient of $1\over\epsilon$ in the corresponding charge
is computed as
$\Delta_{\Phi}S=\int d\sigma \Phi(X(\sigma))$.  Let us
consider $\Delta_{\Phi}S$ as a small perturbation of the
action of the Neumann system. In other words, take very small $\Phi$
and add $\Delta_{\Phi}S$ to the action. Then the value
of $\Delta_{\Phi}S$ on a given periodic trajectory of the Neumann system
can be computed as the first order  correction to the action
on the periodic trajectory of the perturbed system. But the first order
correction to the action is actually zero. Indeed, $\Phi^{I\bar{J}}$
and $\tilde{\Phi}^{I\bar{J}}$ are Hermitean matrices, therefore 
the potential in the perturbed system would correspond to
 a pair of Hermitean matrices
$\mbox{diag}(A)+\Phi$,  $\mbox{diag}(\tilde{A})+\tilde{\Phi}$. 
We can diagonalize them by  unitary transformations.
We  get the  Neumann system with the new 
coefficients
$A',\tilde{A}'$. But the differences $A'-A$ and $\tilde{A}'-\tilde{A}$
are of the second order in $\Phi$. Therefore the first order
correction to the action is zero. (See Appendix C of \cite{AFRT}
for an alternative proof.)

But if the contour is arbitrary, not a solution to the nondegenerate
Neumann equation, then the non-diagonal charges may be nonzero.
In general we will get a pair of Hermitean matrices of charges
\begin{equation}
q_{I\bar{J}}=\int d\sigma \; Y_I\bar{Y}_{\bar{J}},\;\;\;\;
\tilde{q}_{I\bar{J}}=\int d\sigma \; Y_I\bar{Y}_{\bar{J}}
\end{equation}
for the string solution which does not minimize the $Q_V$
for given diagonal charges $Q_0,\ldots, Q_5$.

\subsection{Exact worldsheets.}
We have argued that the string solutions can be obtained
as perturbations near the degenerate worldsheet
with the small parameter $\epsilon$. In \cite{AFRT} the 
solutions of the Neumann system were used to construct the
exact worldsheet, rather than an expansion in a small parameter.
More general Neumann system considered here  can also be 
used to build exact  string worldsheets.  
Consider the following ansatz:
\begin{equation}
Y_I(\sigma,t)=e^{iw_I t} Y_I(\sigma),\;\;\;
Z_I(\sigma,t)=e^{i\tilde{w}_I t} Z_I(\sigma)
\end{equation}
where $Y_I(\sigma)$ and $Z_I(\sigma)$ solve the Neumann system
equations of motion: 
$$\partial_{\sigma}^2 Y_I+ Y_I \sum s_J|\partial_{\sigma}Y_J|^2=
-w_I^2 Y_I+ Y_I \sum s_J w_J^2 |Y_J^2|$$
and the analogous equation for $Z$. The functions
$Y_I(\sigma)$ and $Z_I(\sigma)$  should be periodic modulo 
\begin{equation}\label{PeriodicModulo}
(Y_I,Z_I)\to (e^{iw_I\varphi} Y_I, e^{i \tilde{w}_I\varphi}Z_I)
\end{equation}
We have $$(\partial_{t}x,\partial_{t}x)=
\sum s_Iw_I^2|Y_I|^2-\sum \tilde{w}_I^2 |Z_I|^2$$ 
$$(\partial_{\sigma}x,\partial_{\sigma}x)=\sum s_I 
|\partial_{\sigma}Y_I|^2-\sum |\partial_{\sigma}Z_I|^2$$
Let us restrict $w_I,\tilde{w}_I$ to satisfy:
\begin{equation}\label{FiniteConstraint}
\sum s_I w_I  M_I=\sum \tilde{w}_I \tilde{M}_I
\end{equation}
This constraint implies that $(\partial_{\sigma}x,\partial_{t}x)=0$.
The trace of the second fundamental form in the $AdS_5$ direction is:
\begin{equation}
{-w_I^2 Y_I+ Y_I \sum s_J w_J^2 |Y_J^2|
\over \sum s_Iw_I^2|Y_I|^2-\sum \tilde{w}_I^2 |Z_I|^2}
+{\partial_{\sigma}^2 Y_I+ Y_I \sum s_J|\partial_{\sigma}Y_J|^2
\over \sum s_I|\partial_{\sigma}Y_I|^2-\sum |\partial_{\sigma}Z_I|^2}
\end{equation}
and the analogous expression 
in the $S^5$ direction. It is zero for the solutions which have
zero energy:
\begin{equation}
\sum s_I|\partial_{\sigma}Y_I|^2-\sum |\partial_{\sigma}Z_I|^2+
\sum s_Iw_I^2|Y_I|^2-\sum \tilde{w}_I^2 |Z_I|^2=0
\end{equation}
which means that the ansatz solves the equations for the extremal surface.
It is not true that 
 (\ref{PeriodicModulo}) with $\sigma$-dependent $\varphi$
is a symmetry of the action. The Neumann system with the constraint
(\ref{FiniteConstraint}) has nine degrees of freedom.
Gauge symmstry exists only in
the  limit $\kappa\to\infty$ when the constraint becomes
$\sum s_I M_I=\sum \tilde{M}_I$.  

The ultrarelativistic limit corresponds to $w_I^2=A_I+\kappa^2$, 
$\tilde{w}_I^2=\tilde{A}_I+\kappa^2$ with $\kappa^2\to \infty$. 
The parameter $\epsilon$ is of the order $1\over\kappa$, and
$t\simeq \tau/\kappa$.
There is a subtlety with these exact worldsheet solutions;
we do not know if the solutions with different $\kappa$
belong to the same $\lambda$-family.

\subsection{Integrability of the Neumann system (a very brief review.)}
Here we will review the basic properties of the Neumann system, following
mostly \cite{BabelonTalon}.
Consider the dynamical system with the Lagrangian
\begin{equation}
L={1\over 2}\sum\limits_{\mu=1}^N
 s_{\mu}\left((\dot{x}_{\mu})^2 - a_{\mu} x_{\mu}^2\right)
\end{equation}
where $s_{\mu}=\pm 1$ and $\sum s_{\mu} x_{\mu}^2=1$. 
There are $N-1$ independent integrals of motion:
\begin{eqnarray}
F_{\mu}=x_{\mu}^2+
\sum\limits_{\nu\neq \mu} s_{\nu}{(x_{\mu} \dot{x}_{\nu} - 
x_{\nu} \dot{x}_{\mu})^2
\over a_{\mu}-a_{\nu}}\\[5pt]
\sum s_{\mu} F_{\mu}=1
\end{eqnarray}
Introduce the parametrization:
\begin{equation}
x_{\mu}^2=s_{\mu}{\prod_{\nu} (t_{\nu}-a_{\mu})\over 
\prod_{\lambda\neq \mu}(a_{\lambda}-a_{\mu})}
\end{equation}
Using the identities:
\begin{eqnarray}
\sum_{\nu} {s_{\nu} x_{\nu}^2\over 
t_{\rho}-a_{\nu}}=0\\[5pt]
\sum_{\nu} {s_{\nu} x_{\nu}^2\over 
(t_{\rho}-a_{\nu})^2}={\prod_{\sigma\neq\rho}(t_{\rho}-t_{\sigma})\over
\prod_{\nu}(t_{\rho}-a_{\nu})}
\end{eqnarray}
we can rewrite the conserved quantities in $t$-coordinates:
\begin{equation}
F_{\mu}(t,\dot{t})=
x_{\mu}^2\left(1-{1\over 4}\sum_{\rho}
{\dot{t}_{\rho}^{\;2}\over t_{\rho}-a_{\mu} }
{\prod_{\sigma\neq\rho}(t_{\rho}-t_{\sigma})\over
\prod_{\nu}(t_{\rho}-a_{\nu})}\right)
\end{equation}
The trajectories are determined by $F_{\mu}=$const.
They depend on $N-1$ constants $\{b_1,\ldots,b_{N-1}\}$:
\begin{equation}
{d{t}_{\mu}\over d\sigma}
=2\varepsilon_{\mu}{\sqrt{\prod_{\kappa} (t_{\mu}-a_{\kappa})
\prod_{\lambda} (t_{\mu}-b_{\lambda})}
\over 
\prod_{\nu\neq \mu} (t_{\mu}-t_{\nu})}
\end{equation}
Here $\varepsilon_{\mu}=\pm 1$. On these trajectories
$F_{\mu}={\prod_{\lambda}(a_{\mu}-b_{\lambda})\over
\prod_{\nu\neq\mu} (a_{\mu}-a_{\nu})}$.

We are interested in the special case of the Neumann problem, when $N$ 
is even and  pairs of $a_{\mu}$ coincide. 
Consider  the limit $a_{2I}\to a_{2I-1}=A_I$  
with  $t_{{N\over 2}+I-1}$ locked between
$a_{2I}$ and $a_{2I-1}$ for $I=1,\ldots, N/2$. For the remaining $t_I$
with $I=1,\ldots, N/2-1$ we get:
\begin{equation}\label{T}
{dt_I\over d\sigma}=2\varepsilon_I
{\sqrt{\prod_k (t_I-b_k)}\over \prod_{K\neq I}(t_I-t_K)}
\end{equation}
The trajectories of $t_I$ do not depend on $A_I$. 
To describe the oscillations of the coordinates trapped between 
$a_{2I}$ and $a_{2I-1}$ we introduce the angles $\theta_I$:
\begin{equation}
\tan\theta_I=\sqrt{a_{2I}-t_{N/2+I-1}\over 
t_{N/2+I-1}-a_{2I-1}}
\end{equation}
Denote $X_I=x_{2I-1}+ix_{2I}$. We have: $X_I=|X_I|e^{i\theta_I}$
and $|X_I|^2={\prod_K(A_I-t_K)\over \prod_{J\neq I}(A_I-A_J)}$.
The  angles are cyclic variables:
\begin{equation}\label{AnglesTheta}
{d\theta_I\over d\sigma}=\varepsilon_{N/2+I-1}
{\sqrt{-\prod_j(A_I-b_j)}\over\prod_J (A_I-t_J)}
\end{equation}
The corresponding momentum is conserved:
\begin{equation}
M_I=|x_I|^2{d\theta_I\over d\sigma}=\varepsilon_{N/2+I-1}
{\sqrt{-\prod_j (A_I-b_j)}\over\prod_{J\neq I}(A_I-A_J)}
\end{equation}
We can use the identity
$
{1\over\prod_J (A-t_J)}=
\sum_J \left[{1\over \prod_{K\neq J}(t_J-t_K)} \right]{1\over A-t_J}
$ and write:
\begin{equation}
\theta_I=\varepsilon_{N/2+I-1}\sqrt{-\prod_j (A_I-b_j)} \sum_J \int
{\varepsilon_J dt_J\over (A_I-t_J)\sqrt{\prod_k (t_J-b_k)}}
\end{equation}
It is convenient to put $\varepsilon_{N/2+I-1}=1$
by replacing
$X_I$ with its complex conjugate if $\varepsilon_{N/2+I-1}=-1$. 

\subsection{The product of two Neumann systems.}
The motion on $(AdS_5\times S^5)/V$ is described by the product
of two Neumann systems. The projection to $AdS_5$ 
is described by the Neumann system with the parameters
$(A_0,A_1,A_2)$ and the action variables $b_1,\ldots, b_5$. The complex
coordinates $(Y_0,Y_1,Y_2)$ are parametrized by 
$t_1,t_2, \theta_0,\theta_1,\theta_2$:
\begin{equation}
Y_I=|Y_I| e^{i\theta_I},\;\;\;\;
|Y_I|^2=s_I{\prod_J (t_J-A_I)\over\prod_{K\neq I} (A_K-A_I)}
\end{equation}
where $s_I$ is a sign: $(s_0,s_1,s_2)=(1,-1,-1)$. 
The projection of the trajectory on $AdS_5$ is described by 
Eqs. (\ref{T}) and (\ref{AnglesTheta}).
The projection to $S^5$ is described by the Neumann system with the
parameters $\tilde{A}_1,\tilde{A}_2,\tilde{A}_3$; the action variables
are $\tilde{b}_1,\ldots,\tilde{b}_5$. The coordinates on the sphere are:
\begin{equation}
Z_I=|Z_I| e^{i\tilde{\theta}_I},\;\;\;\;
|Z_I|^2={\prod_J (\tilde{t}_J-\tilde{A}_I)\over\prod_{K\neq I} 
(\tilde{A}_K-\tilde{A}_I)}
\end{equation}
The projection of the trajectory on $S^5$ is described by
Eqs. (\ref{T}) and (\ref{AnglesTheta}) with 
$(t,\theta,A,b)\to (\tilde{t},\tilde{\theta},\tilde{A},\tilde{b})$.

We are interested in the periodic trajectories, 
with the period $\int d\sigma = 2\pi$. The Neumann system is integrable,
therefore the phase space is foliated by  invariant tori \cite{Arnold}.
The trajectory in the phase space 
 is $\dot{\phi}_a=\omega_a(I)$ where $\phi_a$ are the angle variables
and $I$ are the action 
variables specifying the torus. For the trajectory to be periodic with
the period $2\pi$  the frequencies $\omega_a$ should be integers: 
\begin{equation}\label{PeriodicTorus}
\omega_a(I)=m_a
\end{equation}
These are the equations on the action variables. The number of unknowns
is equal to the number of equations, therefore we expect in general
to have a discrete set of periodic trajectories, corresponding to the
special values of the action variables. We will say that the torus is
periodic if the frequencies are integer. 
Let us see how it works in our case.

To make the formulas more transparent we consider the system
corresponding to $AdS_{N-1}\times S^{N-1}$ keeping in mind that
we are mostly interested in $N=6$.
We are interested in the solutions which satisfy the constraint
$\sum s_I M_I=\sum \tilde{M}_I$ and are periodic
modulo the overall phase. The constraint reads:
\begin{equation}\label{OrthogonalityConstraint}
\sum_{I=0}^{N-1} s_I 
{\sqrt{-\prod_j (A_I-b_j)}\over\prod_{J\neq I}(A_I-A_J)}
=
\sum_{I=1}^N 
{\sqrt{-\prod_j (\tilde{A}_I-\tilde{b}_j)}\over
\prod_{J\neq I}(\tilde{A}_I-\tilde{A}_J)}
\end{equation}
To formulate the periodicity conditions for $t_I$ and $\tilde{t}_I$
we rewrite the equations (\ref{T}) in the following form:
\begin{eqnarray}
&&\sum_{I} {\varepsilon_I 
t_{I}^k dt_{I}\over \sqrt{\prod_{j}(t_I-b_j)}}=0,\;\;\;\;
\mbox{for}\;\;\; k=0,\ldots, {N\over 2}-3 \\[5pt]
&&\sum_{I} {\varepsilon_I 
t_{I}^{N/2-2} dt_{I}\over \sqrt{\prod_{j}(t_I-b_j)}}=
2\; d\sigma
\end{eqnarray}
We can consider $N/2 -1$ pairs $(t_I,\varepsilon_I)$ as
specifying $N/2-1$ points on the curve ${\cal C}$ described
by the equation
$y^2=\prod_{j=1}^{N-1}(t-b_j)$.
Indeed, $t_I$ gives the value of $t$  and $\varepsilon_I$
fixes the sign of $y$. The periodic trajectory then defines
a homology class  $c\in H_1({\cal C}, {\bf Z})$ as the formal
sum of the trajectories of the points $(t_I,\varepsilon_I)$
for $I=1,\ldots,N/2-1$. We have
\begin{eqnarray}\label{PeriodicityForT}
&&\oint_c {\varepsilon  
t^k dt\over \sqrt{\prod_{j}(t-b_j)}}=
\oint_{\tilde{c}} {\tilde{\varepsilon}  
\tilde{t}^k d\tilde{t}\over 
\sqrt{\prod_{j}(\tilde{t}-\tilde{b}_j)}}=0\;\;\;
\mbox{for}\;\; k=0,\ldots, {N\over 2}-3 ,\\[5pt]
&&\oint_c {\varepsilon
t^{N/2-2} dt\over \sqrt{\prod_{j}(t-b_j)}}=
\oint_{\tilde{c}} {\tilde{\varepsilon}
\tilde{t}^{N/2-2} d\tilde{t}\over 
\sqrt{\prod_{j}(\tilde{t}-\tilde{b}_j)}}=4\pi 
\label{PeriodTwoPi}
\end{eqnarray}
It turns out that the converse is also true \cite{Mumford}. The cycle 
$(c,\tilde{c})\in H_1({\cal C}\times\tilde{\cal C}, {\bf Z})$ 
satisfying (\ref{PeriodicityForT}),
(\ref{PeriodTwoPi})
defines a periodic (modulo cyclic variables) trajectory of the product of 
two Neumann systems with the period  $2\pi$. Therefore 
(\ref{PeriodicityForT}) and (\ref{PeriodTwoPi}) are the periodicity 
conditions for the $t$ variables. 
It is interesting that they do not depend on $A_I$.
We have to also make sure that the cyclic variables $\theta,\tilde{\theta}$
are periodic functions of $\sigma$. The periodicity conditions for 
$\theta,\tilde{\theta}$  read:
\begin{eqnarray}
\sqrt{-\prod_j (A_I-b_j)} \oint_c
{\varepsilon dt\over (A_I-t)\sqrt{\prod_k (t-b_k)}}=
2\pi m_I+\mu
\\[5pt]
\sqrt{-\prod_j (\tilde{A}_I-\tilde{b}_j)}  \oint_{\tilde{c}}
{\varepsilon d\tilde{t}\over 
(\tilde{A}_I-\tilde{t})\sqrt{\prod_k (\tilde{t}-\tilde{b}_k)}}
=2\pi \tilde{m}_I+\mu
\end{eqnarray}
where $\mu\in {\bf R}$ is an undetermined overall phase.

It is convenient to think about not just one Neumann
system, but the whole family of integrable systems with different values
of $A_I$ and $\tilde{A}_I$. 
We have $N/2$ of $A_I$ and $N/2$ of $\tilde{A}_I$, but both
$A$ and $\tilde{A}$ are defined up to a common shift. 
Therefore we have $N-2$ parameters in the integrable Lagrangian.
For the given Lagrangian, we have $N-1$ of $b$ and $N-1$ of $\tilde{b}$,
the total of $2N-2$ parameters specifying the invariant torus.
The total number of parameters is therefore $3N-4$.
Let us count the constraints. We have one constraint
(\ref{OrthogonalityConstraint}) that the trajectory is orthogonal
to $V$, $N/2-1$ for the periodicity of $t_I$, $N/2-1$ for the
periodicity of $\tilde{t}_I$, and $N-1$ for the periodicity
of $\theta_I$ and $\tilde{\theta}_I$ modulo an overall phase $\mu$.
The total number of constraints is therefore $2N-2$. 

Thus we expect that in the space of parameters $A,\tilde{A}$ 
and action variables $b,\tilde{b}$ 
there is the $N-2$ dimensional subspace corresponding to the
periodic trajectories. The dimension of this subspace
coincides with the number of independent
charges $q_A=\int_0^{2\pi} d\sigma |X_A|^2$. Based on this counting of
parameters, we expect to be able to
 adjust the parameters to get a periodic trajectory 
with  prescribed values
of $q_A$. The kinetic part of the action will then give the
one-loop anomalous dimension. 

\subsection{Special solutions.}
The periodic tori may degenerate.  For example suppose
that $b_1\to b_2$ and $t_1$ is trapped between $b_1$ and $b_2$.
Then (\ref{T}) and (\ref{AnglesTheta}) imply that $b_1, b_2$
and $t_1$ decouple from the equations for $t_2,\ldots, t_{N/2-1}$
and $\theta_2,\ldots,\theta_{N/2-1}$. We do not have to impose
the periodicity condition on $t_1$, because in the limit $b_1\to b_2$
we have $t_1=$const. 
In fact the value of $b_1=b_2$
does not enter in  the remaining periodicity conditions (it does
however enter the constraint (\ref{OrthogonalityConstraint}).)
Therefore we have one less parameter (because we impose $b_1=b_2$)
but also one less constraint. This means that we still have the
$N-2$-parameter family of periodic trajectories corresponding to
$b_1=b_2$. We conclude that the space of periodic trajectories
consists of several branches, corresponding to  degenerations
of the periodic tori. 

An interesting special case corresponds to $N-2$ of the parameters
$b$ coinciding pairwise. 
Suppose that $b_{2I}\to b_{2I-1}=B_I$ for $I=1,\ldots,N/2-1$
and use the shift symmetry to put $b_{N-1}=0$. Consider $t_J$ oscillating
between $b_{2J-1}$ and $b_{2J}$. From (\ref{AnglesTheta}) we get the 
periodicity conditions:
\begin{equation}
A_I=(m_I+\mu)^2,\;\;\;\; \tilde{A}_I=(\tilde{m}_I+\mu)^2
\end{equation}
where $m_I$ are integers and $\mu$ is real. The oscillation of $t_J$ 
between $b_{2J-1}$ and $b_{2J}$ is described by the equation:
\begin{equation}\label{AnglesPhi}
{d\phi_I\over d\sigma}=\sqrt{B_I}
\end{equation}
Let us put $B_I=n_I^2$. The $n_I$ does not have to be an integer,
because the coordinate $\phi_I$ degenerates when 
$b_{2J-1}\to b_{2J}$. But if $n_I$ is integer then we can resolve
$B_I$ into a pair $b_{2I-1}\neq b_{2I}$. This means that the periodic
trajectories with integer $n_I$ are at the intersection of the different
branches. Independently of whether or not $n_I$ are integer we can 
explicitly write down the corresponding periodic trajectories. 
The absolute values of $Z_I$ and $Y_I$ are constant, and the
phases depend on $\sigma$ linearly:
\begin{equation}\label{LinearPhases}
Y_I(\sigma)=e^{ i (m_I+\mu)\sigma} |Y_I|,\;\;\;\;\;
Z_I(\sigma)=e^{ i (\tilde{m}_I+\mu)\sigma} |Z_I|
\end{equation}
\begin{eqnarray}\label{TotallyDegenerateOne}
&&|Y_I|^2={1\over 2\pi}q_I=
s_I{\prod_J [(m_I+\mu)^2-n_J^2]\over \prod_{J\neq I}
[(m_I+\mu)^2-(m_J+\mu)^2] }\\[5pt]
&&|Z_I|^2={1\over 2\pi}\tilde{q}_I=
{\prod_J [(\tilde{m}_I+\mu)^2-\tilde{n}_J^2]\over \prod_{J\neq I}
[(\tilde{m}_I+\mu)^2-(\tilde{m}_J+\mu)^2] }\label{TotallyDegenerateTwo}
\end{eqnarray}
The momenta are:
\begin{eqnarray}
&&M_I=s_I(m_I+\mu) {\prod_J [(m_I+\mu)^2-n_J^2]\over \prod_{J\neq I}
[(m_I+\mu)^2-(m_J+\mu)^2] }\\[5pt]
&&\tilde{M}_I=(\tilde{m}_I+\mu) 
{\prod_J [(\tilde{m}_I+\mu)^2-\tilde{n}_J^2]\over \prod_{J\neq I}
[(\tilde{m}_I+\mu)^2-(\tilde{m}_J+\mu)^2] }
\end{eqnarray}
The action is:
\begin{equation}
S=\sum (m_I+\mu)^2 -\sum n_J^2+ 
\sum (\tilde{m}_I+\mu)^2 -\sum \tilde{n}_J^2
\end{equation}
and the constraint $\sum s_I M_I=\sum\tilde{M}_I$ reads:
\begin{eqnarray}
&&-(m_0+\mu)|Y_0|^2+(m_1+\mu)|Y_1|^2+(m_2+\mu)|Y_2|^2+\\[5pt]
&&+(\tilde{m}_1+\mu)|Z_1|^2 +(\tilde{m}_2+\mu)|Z_2|^2
+(\tilde{m}_3+\mu)|Z_3|^2=0
\end{eqnarray}
One can see that $\mu$ drops out of this constraint (enters only through
$|Z_I|^2$ and $|Y_I|^2$) because $\sum s_I|Y_I|^2=\sum |Z_I|^2$.
Therefore the constraint actually imposes a restriction on the
possible values of the R-charges and the spins:
\begin{equation}
\sum s_I m_I q_I =\sum \tilde{m}_I \tilde{q}_I
\end{equation} 
This restriction on charges and spins
 is a feature of the "totally degenerate"
periodic trajectories. For generic
$n_I,\tilde{n}_I$ the small deformations of these totally degenerate
 trajectories correspond to varying $n_I,\tilde{n}_I$ in
(\ref{TotallyDegenerateOne}), 
(\ref{TotallyDegenerateTwo}). 
But if some of the $n_I, \tilde{n}_I$
 are integer, then there are additional deformations corresponding
to splitting the corresponding pair of coinciding $b_j$. It would be
interesting to understand what is special about these "integer"
values of the R-charge and the spin in the dual gauge theory.

Another special case is when $(b_1,b_2,b_3)\to (A_0,A_1,A_2)$
and $(\tilde{b}_1,\tilde{b}_2,\tilde{b}_3)\to 
(\tilde{A}_1,\tilde{A}_2,\tilde{A}_3)$. In this case all the angles
$\theta_I$, $\tilde{\theta}_I$ are frozen. It is the same
as restricting $Y_I$ and $Z_I$ to be real. 
This case was considered in \cite{AFRT}.

\section*{Acknowledgements.} 
I want to thank P.~Bozhilov, M.~Kruczenski, S.~Moriyama and A.~Tseytlin
for interesting discussions.  This
research was supported in part by the National Science Foundation under
Grant No. PHY99-07949 and in part
by the RFBR Grant No. 00-02-116477 and in part by the 
Russian Grant for the support of the scientific schools
No. 00-15-96557.

\end{document}